\documentclass[
reprint,
superscriptaddress,
amsmath,amssymb,
aps,
prb,
longbibliography
]{revtex4-1}

\usepackage{graphicx,epsf}
\usepackage{float}
\usepackage{xcolor}
\usepackage{bm}

\usepackage{cancel}

\begin{document}

\title{
Trivial and topological bound states  
in bilayer graphene quantum dots and rings
}

\author{Nassima Benchtaber}
\affiliation{Institute for Cross-Disciplinary Physics and Complex Systems IFISC 
(CSIC-UIB), E-07122 Palma, Spain} 
\author{David S\'anchez}
\affiliation{Institute for Cross-Disciplinary Physics and Complex Systems IFISC 
(CSIC-UIB), E-07122 Palma, Spain} 
\affiliation{Department of Physics, University of the Balearic Islands, 
E-07122 Palma, Spain}
\author{Lloren\c{c} Serra}
\affiliation{Institute for Cross-Disciplinary Physics and Complex Systems IFISC 
(CSIC-UIB), E-07122 Palma, Spain} 
\affiliation{Department of Physics, University of the Balearic Islands, 
E-07122 Palma, Spain}

\begin{abstract}
We discuss and compare two different types of confinement in bilayer graphene 
by top and bottom gating with symmetrical microelectrodes. Trivial 
confinement corresponds to the same polarity of all top gates,
which is 
opposed to that of all bottom ones. 
Topological confinement requires  the  
polarity of part of the top-bottom pairs of gates to be reversed. 
We show that the main qualitative difference between 
trivial and topological bound states 
manifests itself in the magnetic field dependence.
We illustrate our finding with an explicit calculation of the energy spectrum
for quantum dots and rings.
Trivial confinement shows bunching of levels into degenerate Landau
bands, with a non-centered gap, while topological confinement 
shows no field-induced gap and a sequence of state branches 
always crossing zero-energy. 
\end{abstract}

\maketitle

\section{Introduction}
\label{Intro}

The topological nature of bound states is a research question of ongoing
interest~\cite{fu08,kit12}.
Partly due to the many applications that these states might have for quantum
computation~\cite{nay08}, it is crucial to propose reliable tests that help differentiate between
bound states arising from trivial confinement and those arising from topological
confinement. The latter show more robustness against weak disorder
but it is not easy to detect their presence~\cite{sch20}, which is a prerequisite for their subsequent
manipulation and measurement in topological quantum information tasks.

Bilayer graphene (BLG)~\cite{Mcan13} in a Bernal stacking structure
is a especially suitable platform for creating
solid-state electronic qubits due to its long decoherence times~\cite{Trau07}.
Confinement in BLG is achieved by means of a pair of gates
applied to the graphene sheets. In this manner, an energy gap opens
around the Fermi level, which is the building block for tunnel barriers~\cite{Over18,Kraf18}.
The only requirement is that the potential applied to a sheet has the sign
opposite to that of the second sheet.
This mechanism has been proven successfully in the recent years
when serially connecting tunnel barriers~\cite{Eich18,Kurzmann19,Banszerus20,Banszerus21}.
This confinement is dubbed
trivial because the electronic wave functions vanish asymptotically
deep in the barriers. 
Calculations of trivial dots and rings can be 
found, e.g., in Refs.\ 
\onlinecite{Pereira07,Recher09,Zarenia09,Pereira09,Zarenia10,Zarenia10b}.

A totally different confinement mechanism 
arises when the gate potential changes sign on the same sheet
(with the corresponding sign reversal on the second sheet)~\cite{Mar08}. This can be
done with inhomogeneous potentials such as those occurring across
a domain wall. As a consequence, there appear kink electronic states (two per valley)
bounded along the wall. They are topological because their motion is along
the edge and exhibit valley-momentum locking~\cite{Mar08,Zarenia11,Ben21}. When the wall forms a closed
loop, one can create quantum dots and rings supporting topological bound
states~\cite{xavier10,Ben21b}. In this paper, we show that these states could be unambiguously 
detected by examining their behavior with an external magnetic field, which
allows us to distinguish between both (trival and topological) binding mechanisms.

\begin{figure}[b]
\begin{center}
\includegraphics[width=0.5\textwidth, trim = 2cm 0.5cm 2cm 0.5cm, clip]{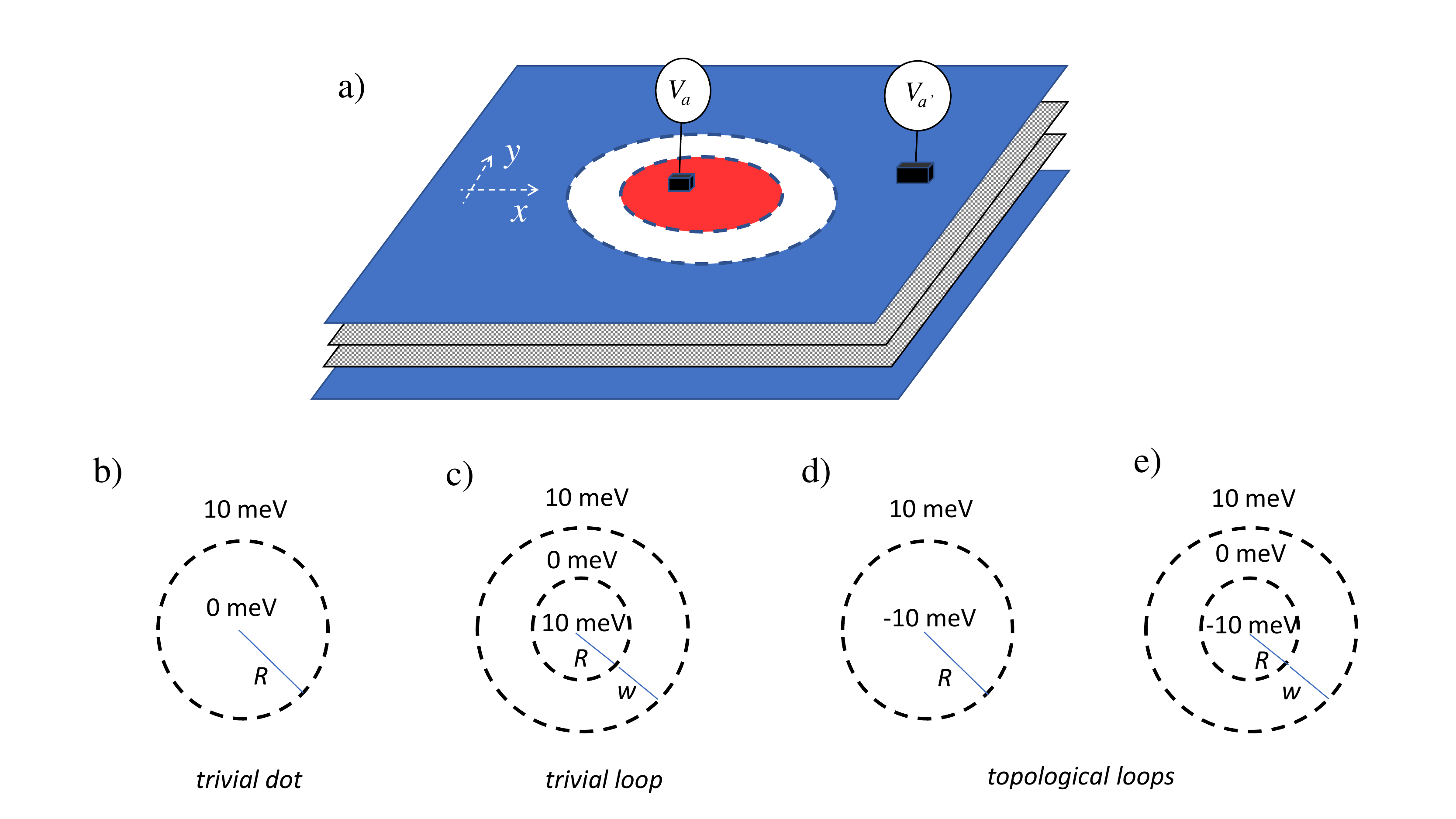}
\end{center}
 \caption{a) Generic distribution of top-bottom gate pairs 
inducing the formation of bound states in BLG. Only the top gates are shown (red and blue), 
with a similar ditribution of bottom gates hidden under the two graphene layers (gray planes).
The potentials $V_a$, $V'_a$ are reversed in the bottom gates.
b)-e) Selected configurations for bound states in  
BLG nanostructures induced by the indicated
gate potentials: trivial dot of radius $R$ (b), trivial 
ring of radius $R$ and width $w$ (c), and 
topological rings of radius $R$ with vanishing width (d) and 
finite width $w$ (e).  
}
\label{Fig1}
\end{figure}

Figure \ref{Fig1} shows a sketch of a BLG system with  
top-bottom gating defining circular nanostructures. 
The distribution of gates is symmetric, i.e., the same for top and bottom 
sheets while the applied potentials $V_a$ on the top gates is sign 
reversed with respect to the bottom gates. This way,  
a potential difference is created between the two graphene 
layers sandwiched  by a given pair of top-bottom gates.
Due to the proximity of both graphene layers, much smaller than 
the gates separation, the effective interlayer field will be  
smaller than the inter-gate field, but it may be tuned in a 
proportional way. 

We call {\em trivial confinement} the case when all gates on a given side
of the BLG planes,
top or bottom, have the same sign. 
This corresponds to an electric field always in the same direction
across the BLG planes, say upwards, 
confinement being caused by the preference of 
electrons to attach to the 
regions of low or vanishing electric field. 
This is indicated by  
the white region in Fig.\ \ref{Fig1}a, and in the 0 meV region of
Fig.\ \ref{Fig1}b
(trivial dot) and Fig.\ \ref{Fig1}c (trivial ring).

{\em Topological confinement} requires gates on a given side, top or bottom, 
to have potentials of different signs. This way, the BLG interlayer field changes 
direction creating  a topological domain wall able to bind electrons.
Figure \ref{Fig1}d
sketches the situation of a topological circular ring.
It would correspond to the top gates 
of blue and red color 
in Fig.\ \ref{Fig1}a having potentials 10 meV and -10 meV, respectively, with a sign change in the white region.
We 
also
stress that, differently to the trivial confinement, the topological 
confinement does not require 
an extended region of vanishing field; even an abrupt domain wall
Fig.\ \ref{Fig1}d
creates bound states. This is already indicating the predominantly  
1d character of topological states in BLG, as opposed to the 
predominantly 2d character 
of the trivial confinement. As shown below, this results 
in conspicuous physical differences regarding the spectrum 
dependence on magnetic field.

We can now present the main finding of this work.
In the presence of a  perpendicular magnetic field, trivial confinement 
spectra show a bunching of levels indicating the emergence of the
2d physics of Landau levels in BLG. The larger the 2d region of trivial states,
the smaller the field at which Landau-level bunching is observed. At zero field
there is a finite-size spectrum discretization that evolves as a function of the field 
into the mentioned level bunching and the formation of a gap 
for energies in $[0,\sqrt{2}\hbar\omega_c]$, where
$\omega_c$ is the BLG intrinsic cyclotron energy defined below.
Quite differently, topological structures like those in Fig.\ \ref{Fig1}d,e 
do not show the formation of a gap with magnetic field, but 
a periodic repetition of continuum branches crossing zero energy.
Below, we provide quantitative comparisons of the magnetic energy spectra
of BLG trivial and topological confinements, emphasizing the differences between both 
types. These comparisons are based on numerical calculations that use radial grids with small spacings, allowing high precision for circular dots and rings. In some cases we also consider 2d grid calculations, only to confirm the physical scenario for systems departing from circular symmetry.

\section{Model}

We consider a 2d $(xy)$ continuum model for the low-energy excitations of BLG, already used in our previous works,\cite{Ben21,Ben21b} and also by many other 
authors (see Refs.\ \onlinecite{Mcan13,rozhkov16} for reviews).
The Hamiltonian reads
\begin{eqnarray}
 H &=& v_F \left(p_x - \hbar  \frac{y}{2l_z^2}\right) \tau_z \sigma_x
 + v_F\, 
 \left( p_y+ \hbar  \frac{x}{2l_z^2}\right) \sigma_y \nonumber\\
 &+& \frac{t}{2}\, \left(\,\lambda_x \sigma_x +\lambda_y\sigma_y\,\right) 
 + V_a(x,y)\, \lambda_z\; ,
\label{eq1}
 \end{eqnarray}
 Here,  
$\hbar v_F= 660\, {\rm meV}\,{\rm nm}$  and $t=380\,{\rm meV}$ are the 
Fermi velocity and interlayer coupling, respectively, which are BLG intrinsic 
parameters. The sublattice, layer and valley
two-fold discrete degrees of freedom of BLG are  
represented by the $\sigma_{x,y,z}$, $\lambda_{x,y,z}$ and $\tau_{x,y,z}$ sets of Pauli matrices, respectively. The influence of a vertical magnetic field $B$ is
included by means of the magnetic length parameter 
$l_z=\sqrt{\hbar/eB}$, in a symmetric gauge affecting the $p_x$ and $p_y$ operators.
Notice that $H$ contains only linear momentum terms, a characteristic of Dirac or relativistic-like Hamiltonians. 

A remarkable property of BLG is that 
confinement to nanostructures can be achieved with the Hamiltonian of Eq.\ (\ref{eq1})
by space modulation of the layer-asymmetry potential $V_a(x,y)$. This is a potential imbalance between the two graphene layers that can be tuned by top and bottom gating, as sketched in Fig.\ \ref{Fig1}. Physically, the electrons have a preference 
to stay in regions where this potential imbalance is lower and this can be exploited 
to confine them in nanostructures whose shape is controlled by the shape of the gates.
A simplest geometry of confinement that has attracted much attention is the circularly symmetric shape, both as quantum dots and rings. Noncircular shapes 
have been discussed in Ref. \onlinecite{Ben21b}.

In this work we consider a layer-asymmetry potential of circular shape, parameterized as
\begin{eqnarray}
V_a(r) &=&
V_a^{({\it in})}
\frac{1}{1+e^{(r-R)/s}}
\nonumber\\
&+&
V_a^{({\it out})}
\left(1 -
\frac{1}{1+e^{(r-R-w)/s}}
\right)
\; ,
\label{eq2}
\end{eqnarray}
where $V_a^{({\it in})}$ represents an inner saturation value for $r<R$ and
$V_a^{({\it out})}$ an outer saturation value for $r>R+w$. The potential $V_a(r)$ is vanishing for $r\in[R,R+w]$ and $s$ is a small diffusivity 
suggested by realistic electrostatic simulations of straight
kinks,\cite{Li16} which is also convenient
for numerical stability.
Appropriately choosing parameters $R$, $w$ and $V_a^{({\it in,out})}$ it is possible 
to model the different types of confinements sketched in Fig.\ \ref{Fig1}.
These radial potentials can be created by disk-like microelectrodes, as suggested 
in Fig.\ \ref{Fig1}.

As mentioned in Sec.\ \ref{Intro}, the aim of this work is to compare two qualitatively different types of confinement in BLG: trivial confinement corresponding
to saturation potentials of the same sign, 
$
{\rm sgn}({V_a^{({\it in})}})
=
{\rm sgn}({V_a^{({\it out})}})
$,
and topological confinement corresponding to saturation potentials of different
signs, 
$
{\rm sgn}({V_a^{({\it in})}})
\neq
{\rm sgn}({V_a^{({\it out})}})
$.
Our parameterization allows a flexible modelling of trivial dots and rings. 
It also allows us to describe topological rings of zero or finite widths. 
Previous works have investigated 
trivial dots and rings as well as topological rings, but
the latter only with $w=0$.\cite{xavier10,Ben21b} 
Here we will also explore
the case of a topological ring with a finite $w$, where potential $V_a(r)$ 
vanishes and the electrons are essentially free to move.

In presence of a vertical magnetic field electron states in bulk BLG, with $V_a(r)=0$,
are characterized by the emergence of discrete Landau levels with energies\cite{Mcan13}
\begin{equation}
\left\{
\begin{array}{rcl} 
 E_0 &=& 0\; ,\\
 E_1 &=& 0\; ,\\
 E_{\ell,\pm} &=& \pm \hbar\omega_c \sqrt{\ell(\ell-1)}\;,\;\; \ell = 2,3,4\ldots
\end{array}
\right.
\label{eq3}
\end{equation}
There exists a two-fold degenerate Landau level at zero energy and a 
sequence of field-dispersing levels at both positive and negative energies. 
The cyclotron frequency in Eq.\ (\ref{eq3}) is 
$\omega_c = eB/m_0$, with a mass parameter given by the BLG 
intrinsic parameters, $m_0=t/2v_F^2$.
The spectrum of Eq.\ (\ref{eq3}) is degenerate for both valleys.

In finite structures like those in Fig.\ \ref{Fig1} we can expect the emergence of Landau levels for high enough fields provided that the system contains a 2d-like region, i.e., a region with $V_a=0$ where locally electron motion is free and the system resembles bulk BLG. We will show below that, indeed, our calculations indicate Landau level formation in trivial dots (Fig.\ref{Fig1}b), 
trivial rings (Fig.\ref{Fig1}c) and topological rings of finite width (Fig.\ref{Fig1}e), but not in topological rings of vanishing width
(Fig.\ref{Fig1}d). The latter only contain 1d-like loop states, 
whose energies show
$B$-periodic repetitions of linearly dispersing branches reflecting the Aharonov-Bohm periodicities in the flux piercing the loop.

A practical advantage of the circular symmetry is that we can define subspaces of
fixed angular momentum $m$, performing independent diagonalizations in each $m$ subspace. Notice, first, that valley subspaces are always independent in
the Hamiltonian of Eq.\ (\ref{eq1}), irrespectively of the spatial circular or 
noncircular symmetry. We can then assume $\tau_z\equiv 1$, with the reversed valley $\tau_z\equiv-1$ eigenvalues being given by symmetry arguments reversing the energy signs of the $\tau_z\equiv1$ eigenvalues. In the remaining sublattice and layer subspaces, the spatial wave function for angular momentum $m$ can be written 
as a 4 component spinor
\begin{equation}
\left(
\begin{array}{c}
e^{i(m-1)\theta} C_1(r) \\
e^{im\theta} C_2(r) \\
e^{im\theta} C_3(r) \\
e^{i(m+1)\theta} C_4(r) \\
\end{array}
\right)
\label{eq4}
\end{equation}
with $(r,\theta)$ the polar coordinates. 

It can be shown that, with the spinor wave function of Eq.\ (\ref{eq4}) and
Hamiltonian $H$ of Eq.\ (\ref{eq1}), one can fully remove the $\theta$-dependencies 
and diagonalize a purely radial Hamiltonian $H_m(r,p_r)$ for each angular momentum
$m$, 
\begin{eqnarray}
H_m(r,p_r) &=&
v_F\, p_r\, \sigma_x \nonumber\\
&+&
\hbar v_F 
\left(
\frac{m}{r}+\frac{r}{2l_z^2}
\right)\, \sigma_y
-
\hbar v_F 
\frac{1}{2r}
\, \sigma_y\lambda_z \nonumber\\
&+&
\frac{t}{2}(\lambda_x\sigma_x+\lambda_y\sigma_y)
+
V_a(r)\, \lambda_z\; ,
\label{eq5}
\end{eqnarray}
with $p_r=-i\hbar\, d/dr$ the radial momentum.

Hamiltonain $H_m$ can be diagonalized
for a given layer-asymmetry potential $V_a(r)$ using finite differences
in a radial grid and imposing the zero condition at the boundaries. 
Notice that with $m\ne 0$
there is a $1/r$ divergence at the origin in Eq.\ (\ref{eq5}), which is 
compensated by the behavior of the wave function. Numerically, this is more 
easily taken into account by including a finite value of $R$ (even if quite small)  as in trivial and topological rings.
An important aspect to bear in mind in the grid diagonalization of Eq.\ (\ref{eq5})
is the possible appearance of spurious solutions due to the known-problem of Fermion doubling for Dirac Hamiltonians. We have carefully considered this, filtering out spurious solutions by defining grid-average wave functions and elliminating 
those solutions whose norm is affected by such grid averaging.\cite{Ben21}
As mentioned in Sec.\ \ref{Intro}, we have also performed in some test cases the diagonalization of the Hamiltonian given by Eq.\ (\ref{eq1}) without separating in 
subspaces of angular momentum. However, this is much more demanding computationally
and we have only checked the agreement of both methods in a few selected cases.

\begin{figure}[t]
\begin{center}
\includegraphics[width=0.45\textwidth, trim = 0.75cm 7cm 2cm 3cm, clip]{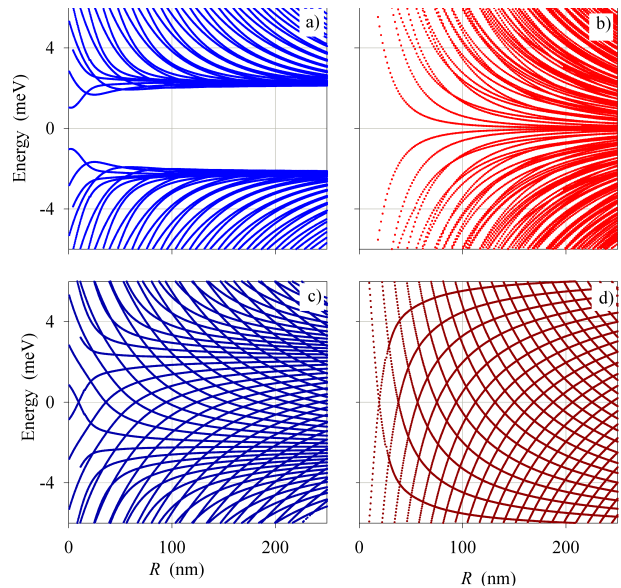}
\end{center}
 \caption{Spectrum of eigenvalues with vanishing magnetic field.
(a) Trivial ring, 
$V_a^{({\it in})}=V_a^{({\it out})}=10\, {\rm meV}$, $w=50\, {\rm nm}$;
(b) trivial dot, 
$V_a^{({\it in})}=0$, $V_a^{({\it out})}=10\, {\rm meV}$, $w=0$; 
(c) topological ring
$V_a^{({\it in})}=-V_a^{({\it out})}=-10\, {\rm meV}$, $w=50\, {\rm nm}$; 
(d) topological ring
$V_a^{({\it in})}=-V_a^{({\it out})}=-10\, {\rm meV}$, $w=0$.
In all cases we used $s=2.5\, {\rm nm}$.}
\label{FigNass}
\end{figure}

\section{Results}

Figure \ref{FigNass} presents selected results for trivial and topological
circular
systems as a function of size, in zero magnetic field. 
Notice that we associate the trivial or topological character to the 
type of confinement and, therefore, this character is shared by the whole low-energy spectrum of eigenstates associated with that particular type of confinement.
A trivial ring 
(Fig.\ \ref{FigNass}a) is characterized by a conspicuous gap in the spectrum around zero energy, due to the energy quantization induced by the finite $w$. In trivial dots or rings with smaller 
$w$'s this energy gap is much reduced or totally closed for large 
values of the radius (Fig.\ \ref{FigNass}b). In sharp contrast, the topological systems present a qualitatively different behavior (Fig.\ \ref{FigNass}cd). Intersecting energy branches of positive and negative slopes, 
always crossing zero energy are the  main characteristic of the topological systems.  
In $w=0$ topological loops
(Fig.\ \ref{FigNass}d)
the pattern of crossings is very regular and, as discussed in Ref.\ \onlinecite{Ben21b}, can be explained with a  quantization rule for 1d closed orbits, similar to the Bohr-Sommerfeld one.  
The topological ring of finite width (Fig.\ \ref{FigNass}c) presents a remarkable behavior, simultaneously showing the zero energy intersecting branches and also a merging of horizontal branches at energies $\approx\pm 2\,{\rm meV}$, clearly  reminiscent of the
gap in a trivial ring of the same size (Fig.\ \ref{FigNass}a).

\begin{figure}[t]
\begin{center}
\includegraphics[width=0.45\textwidth, trim = 0.75cm 17cm 0.5cm 2.5cm, clip]{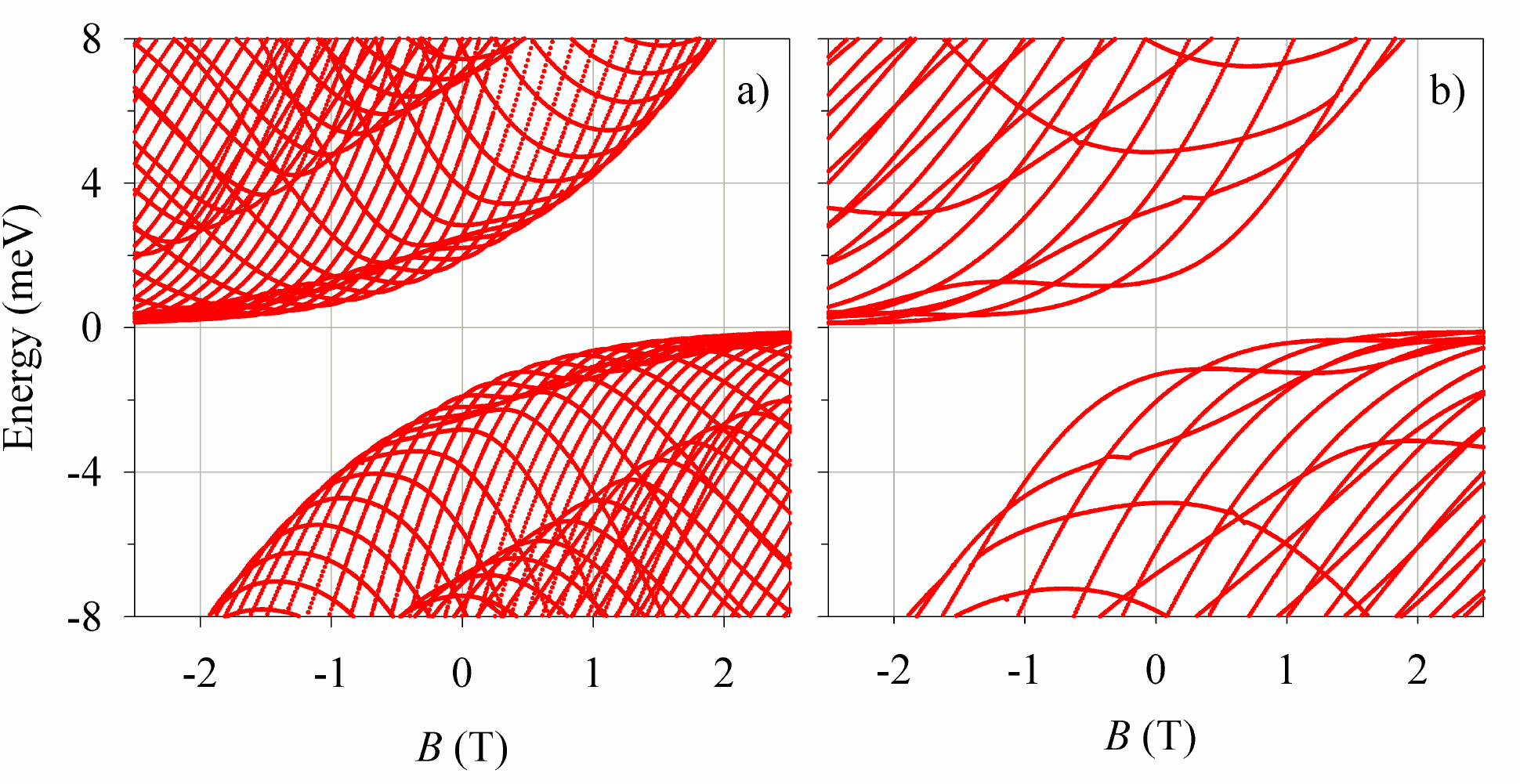}
\end{center}
 \caption{Spectra as a function of  perpendicular magnetic field of trivial rings 
(Fig.\ \ref{Fig1}c) with 
(a) $R=50\, {\rm nm}$, $w=50\, {\rm nm}$;
(b) $R=10\, {\rm nm}$, $w=50\, {\rm nm}$.
Panel (b) is similar to a trivial dot.
We used the same 
$V_a^{({\it in})}$,
$V_a^{({\it out})}$
and  $s$ of Fig.\ \ref{FigNass}.
}
\label{Fig2}
\end{figure}

We consider next the role of a perpendicular magnetic field. 
As discussed above, 
in systems with regions of 2d electronic motion there is a competition of finite size and $B$-field discretization into Landau levels; the latter being eventually dominant for large enough fields. Figures \ref{Fig2}a and \ref{Fig2}b  
show the 
results for trivial rings with a large and a small width $w$, respectively; the latter resembling a quantum dot. Figure \ref{Fig2} is for one valley ($\tau_z=+1$), with the spectrum for the complementary valley ($\tau_z=-1$) being given by reversing the energy signs.
The energy states have an exact symmetry by inversion of both magnetic field and valley. This implies complete valley degeneracy at zero field,
which is the time-reversal-invariant limit.
The most noticeable feature in Fig.\ \ref{Fig2} is the evolution of the energy gap position. Zero energy is the gap center for $B=0$ but it evolves into a
gap edge at large fields. This is a clear indication of Landau band discretization.
For instance, with positive magnetic fields 
the bulk Landau gap $[0,\sqrt{2}\hbar\omega_c]$ at 1.5~T is $[0,7.5\,{\rm meV}]$, in good qualitative agreement with the results of Fig.\ \ref{Fig2}.
This figure also shows how
for the same value of $w$ the spectrum of a ring with a larger $R$ contains more bands and a cleaner merging into Landau levels at large fields. 

\begin{figure}[h]
\begin{center}
\includegraphics[width=0.45\textwidth, trim = 0.75cm 17cm 0.5cm 2.5cm, clip]{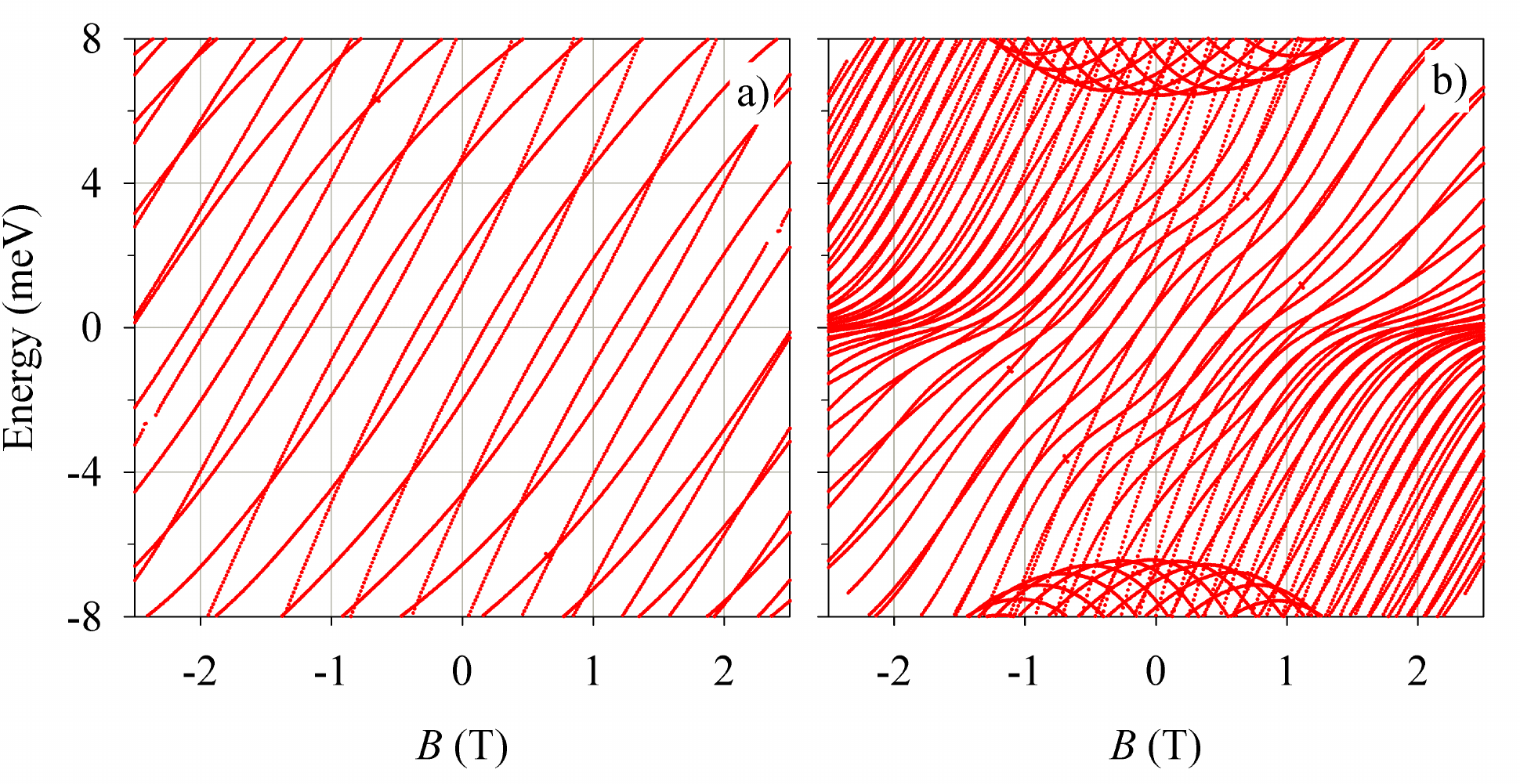}
\end{center}
 \caption{Spectra of topological rings (Fig.\ \ref{Fig1}d,e) with 
(a) $R=50\, {\rm nm}$, $w=0\, {\rm nm}$;
(b) $R=50\, {\rm nm}$, $w=50\, {\rm nm}$.
We used the same  
$V_a^{({\it in})}$,
$V_a^{({\it out})}$
and  $s$ of Fig.\ \ref{FigNass}.
}
\label{Fig3}
\end{figure}

The spectra for topological systems in magnetic field are shown in Fig.\ \ref{Fig3}. As anticipated in Sec.\ \ref{Intro}, the topological loop of zero width (Fig.\ \ref{Fig3}a)
shows no signs of Landau level physics. Instead, its level spectrum is a sequence of almost parallel branches with similar $B$-slopes. Figure \ref{Fig3} is for one valley, the reversed valley having similar branches but with the opposite 
$B$-slopes. 
The case of a topological loop of finite width (Fig.\ \ref{Fig3}b)
is again most remarkable (as in Fig.\ \ref{FigNass}c), showing the 
mentioned almost 
parallel branches and, also, signatures of Landau level discretization.
The latter is hinted by the level bunching around zero energy for large (positive and negative) fields, as well as by the new branches emerging at large (positive and negative) energies in Fig.\ \ref{Fig3}b.

\section{Discussion and conclusions}

We have discussed two types of confinement in BLG nanostructures induced by top and bottom gating. Trivial and topological confinement differ in the potential signs applied to the gates. The regions with zero layer-asymmetry potential correspond to locally free electronic motion. 

Confronting Figs.\ \ref{Fig2} and \ref{Fig3} we observe the sharp differences in the eigenvalue spectra of trivial and topological systems. 
The  $B$-increasing gap (for a given valley), with one gap edge pinned at $E=0$,  
is a characteristic that permits to intrinsically 
differentiate trivial and topological states in BLG systems. Landau level physics requires a 
2d region of vanishing layer-asymmetry potential, where electron motion is locally free. On the contrary, topological rings of vanishing width behave as purely 1d loops and do not show
signatures of Landau level formation. Instead, they manifest $B$-periodicities 
indicating Aharonov-Bohm physics reminiscent of the flux periodicities of
rings built with metals or semiconductor 2d electron gases.

At $B=0$ trivial rings and dots have an energy gap centered around zero energy, larger for the case of rings than for dots. Switching on a magnetic field, this gap evolves into the mentioned non-centered gap of Landau level physics. 
Quite remarkably, topological rings of finite width 
manifest both Aharonov-Bohm periodicities and signatures of Landau level discretization in magnetic field. 

Our numerical estimates suggest that the magnetic spectrum of both confinement types could be detected using today's
experimental techniques, which would represent a significant step toward topological quantum computation in graphene systems.


\acknowledgments
We acknowledge support from AEI (Spain) Grant
No.\ PID2020-117347GB-I00, MINECO/AEI/FEDER Mar\'{\i}a de
Maeztu Program for Units of Excellence MDM2017-0711, 
and GOIB Grant No.\ PDR2020-12.

\bibliography{tritopbib}

\begin{thebibliography}{25}%
\makeatletter
\providecommand \@ifxundefined [1]{%
 \@ifx{#1\undefined}
}%
\providecommand \@ifnum [1]{%
 \ifnum #1\expandafter \@firstoftwo
 \else \expandafter \@secondoftwo
 \fi
}%
\providecommand \@ifx [1]{%
 \ifx #1\expandafter \@firstoftwo
 \else \expandafter \@secondoftwo
 \fi
}%
\providecommand \natexlab [1]{#1}%
\providecommand \enquote  [1]{``#1''}%
\providecommand \bibnamefont  [1]{#1}%
\providecommand \bibfnamefont [1]{#1}%
\providecommand \citenamefont [1]{#1}%
\providecommand \href@noop [0]{\@secondoftwo}%
\providecommand \href [0]{\begingroup \@sanitize@url \@href}%
\providecommand \@href[1]{\@@startlink{#1}\@@href}%
\providecommand \@@href[1]{\endgroup#1\@@endlink}%
\providecommand \@sanitize@url [0]{\catcode `\\12\catcode `\$12\catcode
  `\&12\catcode `\#12\catcode `\^12\catcode `\_12\catcode `\%12\relax}%
\providecommand \@@startlink[1]{}%
\providecommand \@@endlink[0]{}%
\providecommand \url  [0]{\begingroup\@sanitize@url \@url }%
\providecommand \@url [1]{\endgroup\@href {#1}{\urlprefix }}%
\providecommand \urlprefix  [0]{URL }%
\providecommand \Eprint [0]{\href }%
\providecommand \doibase [0]{http://dx.doi.org/}%
\providecommand \selectlanguage [0]{\@gobble}%
\providecommand \bibinfo  [0]{\@secondoftwo}%
\providecommand \bibfield  [0]{\@secondoftwo}%
\providecommand \translation [1]{[#1]}%
\providecommand \BibitemOpen [0]{}%
\providecommand \bibitemStop [0]{}%
\providecommand \bibitemNoStop [0]{.\EOS\space}%
\providecommand \EOS [0]{\spacefactor3000\relax}%
\providecommand \BibitemShut  [1]{\csname bibitem#1\endcsname}%
\let\auto@bib@innerbib\@empty
\bibitem [{\citenamefont {Fu}\ and\ \citenamefont {Kane}(2008)}]{fu08}%
  \BibitemOpen
  \bibfield  {author} {\bibinfo {author} {\bibfnamefont {Liang}\ \bibnamefont
  {Fu}}\ and\ \bibinfo {author} {\bibfnamefont {C.~L.}\ \bibnamefont {Kane}},\
  }\bibfield  {title} {\enquote {\bibinfo {title} {Superconducting proximity
  effect and majorana fermions at the surface of a topological insulator},}\
  }\href {\doibase 10.1103/PhysRevLett.100.096407} {\bibfield  {journal}
  {\bibinfo  {journal} {Phys. Rev. Lett.}\ }\textbf {\bibinfo {volume} {100}},\
  \bibinfo {pages} {096407} (\bibinfo {year} {2008})}\BibitemShut {NoStop}%
\bibitem [{\citenamefont {Kitagawa}\ \emph {et~al.}(2012)\citenamefont
  {Kitagawa}, \citenamefont {Broome}, \citenamefont {Fedrizzi}, \citenamefont
  {Rudner}, \citenamefont {Berg}, \citenamefont {Kassal}, \citenamefont
  {Aspuru-Guzik}, \citenamefont {Demler},\ and\ \citenamefont {White}}]{kit12}%
  \BibitemOpen
  \bibfield  {author} {\bibinfo {author} {\bibfnamefont {Takuya}\ \bibnamefont
  {Kitagawa}}, \bibinfo {author} {\bibfnamefont {Matthew~A.}\ \bibnamefont
  {Broome}}, \bibinfo {author} {\bibfnamefont {Alessandro}\ \bibnamefont
  {Fedrizzi}}, \bibinfo {author} {\bibfnamefont {Mark~S.}\ \bibnamefont
  {Rudner}}, \bibinfo {author} {\bibfnamefont {Erez}\ \bibnamefont {Berg}},
  \bibinfo {author} {\bibfnamefont {Ivan}\ \bibnamefont {Kassal}}, \bibinfo
  {author} {\bibfnamefont {Al{\'a}n}\ \bibnamefont {Aspuru-Guzik}}, \bibinfo
  {author} {\bibfnamefont {Eugene}\ \bibnamefont {Demler}}, \ and\ \bibinfo
  {author} {\bibfnamefont {Andrew~G.}\ \bibnamefont {White}},\ }\bibfield
  {title} {\enquote {\bibinfo {title} {Observation of topologically protected
  bound states in photonic quantum walks},}\ }\href {\doibase
  10.1038/ncomms1872} {\bibfield  {journal} {\bibinfo  {journal} {Nature
  Communications}\ }\textbf {\bibinfo {volume} {3}},\ \bibinfo {pages} {882}
  (\bibinfo {year} {2012})}\BibitemShut {NoStop}%
\bibitem [{\citenamefont {Nayak}\ \emph {et~al.}(2008)\citenamefont {Nayak},
  \citenamefont {Simon}, \citenamefont {Stern}, \citenamefont {Freedman},\ and\
  \citenamefont {Das~Sarma}}]{nay08}%
  \BibitemOpen
  \bibfield  {author} {\bibinfo {author} {\bibfnamefont {Chetan}\ \bibnamefont
  {Nayak}}, \bibinfo {author} {\bibfnamefont {Steven~H.}\ \bibnamefont
  {Simon}}, \bibinfo {author} {\bibfnamefont {Ady}\ \bibnamefont {Stern}},
  \bibinfo {author} {\bibfnamefont {Michael}\ \bibnamefont {Freedman}}, \ and\
  \bibinfo {author} {\bibfnamefont {Sankar}\ \bibnamefont {Das~Sarma}},\
  }\bibfield  {title} {\enquote {\bibinfo {title} {Non-abelian anyons and
  topological quantum computation},}\ }\href {\doibase
  10.1103/RevModPhys.80.1083} {\bibfield  {journal} {\bibinfo  {journal} {Rev.
  Mod. Phys.}\ }\textbf {\bibinfo {volume} {80}},\ \bibinfo {pages}
  {1083--1159} (\bibinfo {year} {2008})}\BibitemShut {NoStop}%
\bibitem [{\citenamefont {Sch\"uler}\ \emph {et~al.}(2020)\citenamefont
  {Sch\"uler}, \citenamefont {De~Giovannini}, \citenamefont {H\"ubener},
  \citenamefont {Rubio}, \citenamefont {Sentef}, \citenamefont {Devereaux},\
  and\ \citenamefont {Werner}}]{sch20}%
  \BibitemOpen
  \bibfield  {author} {\bibinfo {author} {\bibfnamefont {Michael}\ \bibnamefont
  {Sch\"uler}}, \bibinfo {author} {\bibfnamefont {Umberto}\ \bibnamefont
  {De~Giovannini}}, \bibinfo {author} {\bibfnamefont {Hannes}\ \bibnamefont
  {H\"ubener}}, \bibinfo {author} {\bibfnamefont {Angel}\ \bibnamefont
  {Rubio}}, \bibinfo {author} {\bibfnamefont {Michael~A.}\ \bibnamefont
  {Sentef}}, \bibinfo {author} {\bibfnamefont {Thomas~P.}\ \bibnamefont
  {Devereaux}}, \ and\ \bibinfo {author} {\bibfnamefont {Philipp}\ \bibnamefont
  {Werner}},\ }\bibfield  {title} {\enquote {\bibinfo {title} {How circular
  dichroism in time- and angle-resolved photoemission can be used to
  spectroscopically detect transient topological states in graphene},}\ }\href
  {\doibase 10.1103/PhysRevX.10.041013} {\bibfield  {journal} {\bibinfo
  {journal} {Phys. Rev. X}\ }\textbf {\bibinfo {volume} {10}},\ \bibinfo
  {pages} {041013} (\bibinfo {year} {2020})}\BibitemShut {NoStop}%
\bibitem [{\citenamefont {McCann}\ and\ \citenamefont
  {Koshino}(2013)}]{Mcan13}%
  \BibitemOpen
  \bibfield  {author} {\bibinfo {author} {\bibfnamefont {Edward}\ \bibnamefont
  {McCann}}\ and\ \bibinfo {author} {\bibfnamefont {Mikito}\ \bibnamefont
  {Koshino}},\ }\bibfield  {title} {\enquote {\bibinfo {title} {The electronic
  properties of bilayer graphene},}\ }\href {\doibase
  10.1088/0034-4885/76/5/056503} {\bibfield  {journal} {\bibinfo  {journal}
  {Reports on Progress in Physics}\ }\textbf {\bibinfo {volume} {76}},\
  \bibinfo {pages} {056503} (\bibinfo {year} {2013})}\BibitemShut {NoStop}%
\bibitem [{\citenamefont {Trauzettel}\ \emph {et~al.}(2007)\citenamefont
  {Trauzettel}, \citenamefont {Bulaev}, \citenamefont {Loss},\ and\
  \citenamefont {Burkard}}]{Trau07}%
  \BibitemOpen
  \bibfield  {author} {\bibinfo {author} {\bibfnamefont {Bj{\"o}rn}\
  \bibnamefont {Trauzettel}}, \bibinfo {author} {\bibfnamefont {Denis~V.}\
  \bibnamefont {Bulaev}}, \bibinfo {author} {\bibfnamefont {Daniel}\
  \bibnamefont {Loss}}, \ and\ \bibinfo {author} {\bibfnamefont {Guido}\
  \bibnamefont {Burkard}},\ }\bibfield  {title} {\enquote {\bibinfo {title}
  {Spin qubits in graphene quantum dots},}\ }\href@noop {} {\bibfield
  {journal} {\bibinfo  {journal} {Nature Physics}\ }\textbf {\bibinfo {volume}
  {3}},\ \bibinfo {pages} {192--196} (\bibinfo {year} {2007})}\BibitemShut
  {NoStop}%
\bibitem [{\citenamefont {Overweg}\ \emph {et~al.}(2018)\citenamefont
  {Overweg}, \citenamefont {Knothe}, \citenamefont {Fabian}, \citenamefont
  {Linhart}, \citenamefont {Rickhaus}, \citenamefont {Wernli}, \citenamefont
  {Watanabe}, \citenamefont {Taniguchi}, \citenamefont {S\'anchez},
  \citenamefont {Burgd\"orfer}, \citenamefont {Libisch}, \citenamefont
  {Fal{'k}o}, \citenamefont {Ensslin},\ and\ \citenamefont {Ihn}}]{Over18}%
  \BibitemOpen
  \bibfield  {author} {\bibinfo {author} {\bibfnamefont {Hiske}\ \bibnamefont
  {Overweg}}, \bibinfo {author} {\bibfnamefont {Angelika}\ \bibnamefont
  {Knothe}}, \bibinfo {author} {\bibfnamefont {Thomas}\ \bibnamefont {Fabian}},
  \bibinfo {author} {\bibfnamefont {Lukas}\ \bibnamefont {Linhart}}, \bibinfo
  {author} {\bibfnamefont {Peter}\ \bibnamefont {Rickhaus}}, \bibinfo {author}
  {\bibfnamefont {Lucien}\ \bibnamefont {Wernli}}, \bibinfo {author}
  {\bibfnamefont {Kenji}\ \bibnamefont {Watanabe}}, \bibinfo {author}
  {\bibfnamefont {Takashi}\ \bibnamefont {Taniguchi}}, \bibinfo {author}
  {\bibfnamefont {David}\ \bibnamefont {S\'anchez}}, \bibinfo {author}
  {\bibfnamefont {Joachim}\ \bibnamefont {Burgd\"orfer}}, \bibinfo {author}
  {\bibfnamefont {Florian}\ \bibnamefont {Libisch}}, \bibinfo {author}
  {\bibfnamefont {Vladimir~I.}\ \bibnamefont {Fal{'k}o}}, \bibinfo {author}
  {\bibfnamefont {Klaus}\ \bibnamefont {Ensslin}}, \ and\ \bibinfo {author}
  {\bibfnamefont {Thomas}\ \bibnamefont {Ihn}},\ }\bibfield  {title} {\enquote
  {\bibinfo {title} {Topologically nontrivial valley states in bilayer graphene
  quantum point contacts},}\ }\href {\doibase 10.1103/PhysRevLett.121.257702}
  {\bibfield  {journal} {\bibinfo  {journal} {Phys. Rev. Lett.}\ }\textbf
  {\bibinfo {volume} {121}},\ \bibinfo {pages} {257702} (\bibinfo {year}
  {2018})}\BibitemShut {NoStop}%
\bibitem [{\citenamefont {Kraft}\ \emph {et~al.}(2018)\citenamefont {Kraft},
  \citenamefont {Krainov}, \citenamefont {Gall}, \citenamefont {Dmitriev},
  \citenamefont {Krupke}, \citenamefont {Gornyi},\ and\ \citenamefont
  {Danneau}}]{Kraf18}%
  \BibitemOpen
  \bibfield  {author} {\bibinfo {author} {\bibfnamefont {R.}~\bibnamefont
  {Kraft}}, \bibinfo {author} {\bibfnamefont {I.~V.}\ \bibnamefont {Krainov}},
  \bibinfo {author} {\bibfnamefont {V.}~\bibnamefont {Gall}}, \bibinfo {author}
  {\bibfnamefont {A.~P.}\ \bibnamefont {Dmitriev}}, \bibinfo {author}
  {\bibfnamefont {R.}~\bibnamefont {Krupke}}, \bibinfo {author} {\bibfnamefont
  {I.~V.}\ \bibnamefont {Gornyi}}, \ and\ \bibinfo {author} {\bibfnamefont
  {R.}~\bibnamefont {Danneau}},\ }\bibfield  {title} {\enquote {\bibinfo
  {title} {Valley subband splitting in bilayer graphene quantum point
  contacts},}\ }\href {\doibase 10.1103/PhysRevLett.121.257703} {\bibfield
  {journal} {\bibinfo  {journal} {Phys. Rev. Lett.}\ }\textbf {\bibinfo
  {volume} {121}},\ \bibinfo {pages} {257703} (\bibinfo {year}
  {2018})}\BibitemShut {NoStop}%
\bibitem [{\citenamefont {Eich}\ \emph {et~al.}(2018)\citenamefont {Eich},
  \citenamefont {Herman}, \citenamefont {Pisoni}, \citenamefont {Overweg},
  \citenamefont {Kurzmann}, \citenamefont {Lee}, \citenamefont {Rickhaus},
  \citenamefont {Watanabe}, \citenamefont {Taniguchi}, \citenamefont {Sigrist},
  \citenamefont {Ihn},\ and\ \citenamefont {Ensslin}}]{Eich18}%
  \BibitemOpen
  \bibfield  {author} {\bibinfo {author} {\bibfnamefont {Marius}\ \bibnamefont
  {Eich}}, \bibinfo {author} {\bibfnamefont {F.}~\bibnamefont {Herman}},
  \bibinfo {author} {\bibfnamefont {Riccardo}\ \bibnamefont {Pisoni}}, \bibinfo
  {author} {\bibfnamefont {Hiske}\ \bibnamefont {Overweg}}, \bibinfo {author}
  {\bibfnamefont {Annika}\ \bibnamefont {Kurzmann}}, \bibinfo {author}
  {\bibfnamefont {Yongjin}\ \bibnamefont {Lee}}, \bibinfo {author}
  {\bibfnamefont {Peter}\ \bibnamefont {Rickhaus}}, \bibinfo {author}
  {\bibfnamefont {Kenji}\ \bibnamefont {Watanabe}}, \bibinfo {author}
  {\bibfnamefont {Takashi}\ \bibnamefont {Taniguchi}}, \bibinfo {author}
  {\bibfnamefont {Manfred}\ \bibnamefont {Sigrist}}, \bibinfo {author}
  {\bibfnamefont {Thomas}\ \bibnamefont {Ihn}}, \ and\ \bibinfo {author}
  {\bibfnamefont {Klaus}\ \bibnamefont {Ensslin}},\ }\bibfield  {title}
  {\enquote {\bibinfo {title} {Spin and valley states in gate-defined bilayer
  graphene quantum dots},}\ }\href {\doibase 10.1103/PhysRevX.8.031023}
  {\bibfield  {journal} {\bibinfo  {journal} {Phys. Rev. X}\ }\textbf {\bibinfo
  {volume} {8}},\ \bibinfo {pages} {031023} (\bibinfo {year}
  {2018})}\BibitemShut {NoStop}%
\bibitem [{\citenamefont {Kurzmann}\ \emph {et~al.}(2019)\citenamefont
  {Kurzmann}, \citenamefont {Overweg}, \citenamefont {Eich}, \citenamefont
  {Pally}, \citenamefont {Rickhaus}, \citenamefont {Pisoni}, \citenamefont
  {Lee}, \citenamefont {Watanabe}, \citenamefont {Taniguchi}, \citenamefont
  {Ihn},\ and\ \citenamefont {Ensslin}}]{Kurzmann19}%
  \BibitemOpen
  \bibfield  {author} {\bibinfo {author} {\bibfnamefont {Annika}\ \bibnamefont
  {Kurzmann}}, \bibinfo {author} {\bibfnamefont {Hiske}\ \bibnamefont
  {Overweg}}, \bibinfo {author} {\bibfnamefont {Marius}\ \bibnamefont {Eich}},
  \bibinfo {author} {\bibfnamefont {Alessia}\ \bibnamefont {Pally}}, \bibinfo
  {author} {\bibfnamefont {Peter}\ \bibnamefont {Rickhaus}}, \bibinfo {author}
  {\bibfnamefont {Riccardo}\ \bibnamefont {Pisoni}}, \bibinfo {author}
  {\bibfnamefont {Yongjin}\ \bibnamefont {Lee}}, \bibinfo {author}
  {\bibfnamefont {Kenji}\ \bibnamefont {Watanabe}}, \bibinfo {author}
  {\bibfnamefont {Takashi}\ \bibnamefont {Taniguchi}}, \bibinfo {author}
  {\bibfnamefont {Thomas}\ \bibnamefont {Ihn}}, \ and\ \bibinfo {author}
  {\bibfnamefont {Klaus}\ \bibnamefont {Ensslin}},\ }\bibfield  {title}
  {\enquote {\bibinfo {title} {Charge detection in gate-defined bilayer
  graphene quantum dots},}\ }\href {\doibase 10.1021/acs.nanolett.9b01617}
  {\bibfield  {journal} {\bibinfo  {journal} {Nano Letters}\ }\textbf {\bibinfo
  {volume} {19}},\ \bibinfo {pages} {5216--5221} (\bibinfo {year}
  {2019})}\BibitemShut {NoStop}%
\bibitem [{\citenamefont {Banszerus}\ \emph {et~al.}(2020)\citenamefont
  {Banszerus}, \citenamefont {Rothstein}, \citenamefont {Fabian}, \citenamefont
  {M{\"o}ller}, \citenamefont {Icking}, \citenamefont {Trellenkamp},
  \citenamefont {Lentz}, \citenamefont {Neumaier}, \citenamefont {Watanabe},
  \citenamefont {Taniguchi}, \citenamefont {Libisch}, \citenamefont {Volk},\
  and\ \citenamefont {Stampfer}}]{Banszerus20}%
  \BibitemOpen
  \bibfield  {author} {\bibinfo {author} {\bibfnamefont {L.}~\bibnamefont
  {Banszerus}}, \bibinfo {author} {\bibfnamefont {A.}~\bibnamefont
  {Rothstein}}, \bibinfo {author} {\bibfnamefont {T.}~\bibnamefont {Fabian}},
  \bibinfo {author} {\bibfnamefont {S.}~\bibnamefont {M{\"o}ller}}, \bibinfo
  {author} {\bibfnamefont {E.}~\bibnamefont {Icking}}, \bibinfo {author}
  {\bibfnamefont {S.}~\bibnamefont {Trellenkamp}}, \bibinfo {author}
  {\bibfnamefont {F.}~\bibnamefont {Lentz}}, \bibinfo {author} {\bibfnamefont
  {D.}~\bibnamefont {Neumaier}}, \bibinfo {author} {\bibfnamefont
  {K.}~\bibnamefont {Watanabe}}, \bibinfo {author} {\bibfnamefont
  {T.}~\bibnamefont {Taniguchi}}, \bibinfo {author} {\bibfnamefont
  {F.}~\bibnamefont {Libisch}}, \bibinfo {author} {\bibfnamefont
  {C.}~\bibnamefont {Volk}}, \ and\ \bibinfo {author} {\bibfnamefont
  {C.}~\bibnamefont {Stampfer}},\ }\bibfield  {title} {\enquote {\bibinfo
  {title} {Electron hole crossover in gate-controlled bilayer graphene quantum
  dots},}\ }\href {\doibase 10.1021/acs.nanolett.0c03227} {\bibfield  {journal}
  {\bibinfo  {journal} {Nano Letters}\ }\textbf {\bibinfo {volume} {20}},\
  \bibinfo {pages} {7709--7715} (\bibinfo {year} {2020})}\BibitemShut {NoStop}%
\bibitem [{\citenamefont {Banszerus}\ \emph {et~al.}(2021)\citenamefont
  {Banszerus}, \citenamefont {Hecker}, \citenamefont {Icking}, \citenamefont
  {Trellenkamp}, \citenamefont {Lentz}, \citenamefont {Neumaier}, \citenamefont
  {Watanabe}, \citenamefont {Taniguchi}, \citenamefont {Volk},\ and\
  \citenamefont {Stampfer}}]{Banszerus21}%
  \BibitemOpen
  \bibfield  {author} {\bibinfo {author} {\bibfnamefont {L.}~\bibnamefont
  {Banszerus}}, \bibinfo {author} {\bibfnamefont {K.}~\bibnamefont {Hecker}},
  \bibinfo {author} {\bibfnamefont {E.}~\bibnamefont {Icking}}, \bibinfo
  {author} {\bibfnamefont {S.}~\bibnamefont {Trellenkamp}}, \bibinfo {author}
  {\bibfnamefont {F.}~\bibnamefont {Lentz}}, \bibinfo {author} {\bibfnamefont
  {D.}~\bibnamefont {Neumaier}}, \bibinfo {author} {\bibfnamefont
  {K.}~\bibnamefont {Watanabe}}, \bibinfo {author} {\bibfnamefont
  {T.}~\bibnamefont {Taniguchi}}, \bibinfo {author} {\bibfnamefont
  {C.}~\bibnamefont {Volk}}, \ and\ \bibinfo {author} {\bibfnamefont
  {C.}~\bibnamefont {Stampfer}},\ }\bibfield  {title} {\enquote {\bibinfo
  {title} {Pulsed-gate spectroscopy of single-electron spin states in bilayer
  graphene quantum dots},}\ }\href {\doibase 10.1103/PhysRevB.103.L081404}
  {\bibfield  {journal} {\bibinfo  {journal} {Phys. Rev. B}\ }\textbf {\bibinfo
  {volume} {103}},\ \bibinfo {pages} {L081404} (\bibinfo {year}
  {2021})}\BibitemShut {NoStop}%
\bibitem [{\citenamefont {Pereira}\ \emph {et~al.}(2007)\citenamefont
  {Pereira}, \citenamefont {Vasilopoulos},\ and\ \citenamefont
  {Peeters}}]{Pereira07}%
  \BibitemOpen
  \bibfield  {author} {\bibinfo {author} {\bibfnamefont {J.~Milton}\
  \bibnamefont {Pereira}}, \bibinfo {author} {\bibfnamefont {P.}~\bibnamefont
  {Vasilopoulos}}, \ and\ \bibinfo {author} {\bibfnamefont {F.~M.}\
  \bibnamefont {Peeters}},\ }\bibfield  {title} {\enquote {\bibinfo {title}
  {Tunable quantum dots in bilayer graphene},}\ }\href {\doibase
  10.1021/nl062967s} {\bibfield  {journal} {\bibinfo  {journal} {Nano Letters}\
  }\textbf {\bibinfo {volume} {7}},\ \bibinfo {pages} {946--949} (\bibinfo
  {year} {2007})}\BibitemShut {NoStop}%
\bibitem [{\citenamefont {Recher}\ \emph {et~al.}(2009)\citenamefont {Recher},
  \citenamefont {Nilsson}, \citenamefont {Burkard},\ and\ \citenamefont
  {Trauzettel}}]{Recher09}%
  \BibitemOpen
  \bibfield  {author} {\bibinfo {author} {\bibfnamefont {Patrik}\ \bibnamefont
  {Recher}}, \bibinfo {author} {\bibfnamefont {Johan}\ \bibnamefont {Nilsson}},
  \bibinfo {author} {\bibfnamefont {Guido}\ \bibnamefont {Burkard}}, \ and\
  \bibinfo {author} {\bibfnamefont {Bj\"orn}\ \bibnamefont {Trauzettel}},\
  }\bibfield  {title} {\enquote {\bibinfo {title} {Bound states and magnetic
  field induced valley splitting in gate-tunable graphene quantum dots},}\
  }\href {\doibase 10.1103/PhysRevB.79.085407} {\bibfield  {journal} {\bibinfo
  {journal} {Phys. Rev. B}\ }\textbf {\bibinfo {volume} {79}},\ \bibinfo
  {pages} {085407} (\bibinfo {year} {2009})}\BibitemShut {NoStop}%
\bibitem [{\citenamefont {Zarenia}\ \emph {et~al.}(2009)\citenamefont
  {Zarenia}, \citenamefont {Pereira}, \citenamefont {Peeters},\ and\
  \citenamefont {Farias}}]{Zarenia09}%
  \BibitemOpen
  \bibfield  {author} {\bibinfo {author} {\bibfnamefont {M.}~\bibnamefont
  {Zarenia}}, \bibinfo {author} {\bibfnamefont {J.~M.}\ \bibnamefont
  {Pereira}}, \bibinfo {author} {\bibfnamefont {F.~M.}\ \bibnamefont
  {Peeters}}, \ and\ \bibinfo {author} {\bibfnamefont {G.~A.}\ \bibnamefont
  {Farias}},\ }\bibfield  {title} {\enquote {\bibinfo {title}
  {Electrostatically confined quantum rings in bilayer graphene},}\ }\href
  {\doibase 10.1021/nl902302m} {\bibfield  {journal} {\bibinfo  {journal} {Nano
  Letters}\ }\textbf {\bibinfo {volume} {9}},\ \bibinfo {pages} {4088--4092}
  (\bibinfo {year} {2009})}\BibitemShut {NoStop}%
\bibitem [{\citenamefont {Pereira}\ \emph {et~al.}(2009)\citenamefont
  {Pereira}, \citenamefont {Peeters}, \citenamefont {Vasilopoulos},
  \citenamefont {Costa~Filho},\ and\ \citenamefont {Farias}}]{Pereira09}%
  \BibitemOpen
  \bibfield  {author} {\bibinfo {author} {\bibfnamefont {J.~M.}\ \bibnamefont
  {Pereira}}, \bibinfo {author} {\bibfnamefont {F.~M.}\ \bibnamefont
  {Peeters}}, \bibinfo {author} {\bibfnamefont {P.}~\bibnamefont
  {Vasilopoulos}}, \bibinfo {author} {\bibfnamefont {R.~N.}\ \bibnamefont
  {Costa~Filho}}, \ and\ \bibinfo {author} {\bibfnamefont {G.~A.}\ \bibnamefont
  {Farias}},\ }\bibfield  {title} {\enquote {\bibinfo {title} {Landau levels in
  graphene bilayer quantum dots},}\ }\href {\doibase
  10.1103/PhysRevB.79.195403} {\bibfield  {journal} {\bibinfo  {journal} {Phys.
  Rev. B}\ }\textbf {\bibinfo {volume} {79}},\ \bibinfo {pages} {195403}
  (\bibinfo {year} {2009})}\BibitemShut {NoStop}%
\bibitem [{\citenamefont {Zarenia}\ \emph
  {et~al.}(2010{\natexlab{a}})\citenamefont {Zarenia}, \citenamefont {Pereira},
  \citenamefont {Chaves}, \citenamefont {Peeters},\ and\ \citenamefont
  {Farias}}]{Zarenia10}%
  \BibitemOpen
  \bibfield  {author} {\bibinfo {author} {\bibfnamefont {M.}~\bibnamefont
  {Zarenia}}, \bibinfo {author} {\bibfnamefont {J.~Milton}\ \bibnamefont
  {Pereira}}, \bibinfo {author} {\bibfnamefont {A.}~\bibnamefont {Chaves}},
  \bibinfo {author} {\bibfnamefont {F.~M.}\ \bibnamefont {Peeters}}, \ and\
  \bibinfo {author} {\bibfnamefont {G.~A.}\ \bibnamefont {Farias}},\ }\bibfield
   {title} {\enquote {\bibinfo {title} {Simplified model for the energy levels
  of quantum rings in single layer and bilayer graphene},}\ }\href {\doibase
  10.1103/PhysRevB.81.045431} {\bibfield  {journal} {\bibinfo  {journal} {Phys.
  Rev. B}\ }\textbf {\bibinfo {volume} {81}},\ \bibinfo {pages} {045431}
  (\bibinfo {year} {2010}{\natexlab{a}})}\BibitemShut {NoStop}%
\bibitem [{\citenamefont {Zarenia}\ \emph
  {et~al.}(2010{\natexlab{b}})\citenamefont {Zarenia}, \citenamefont {Pereira},
  \citenamefont {Chaves}, \citenamefont {Peeters},\ and\ \citenamefont
  {Farias}}]{Zarenia10b}%
  \BibitemOpen
  \bibfield  {author} {\bibinfo {author} {\bibfnamefont {M.}~\bibnamefont
  {Zarenia}}, \bibinfo {author} {\bibfnamefont {J.~Milton}\ \bibnamefont
  {Pereira}}, \bibinfo {author} {\bibfnamefont {A.}~\bibnamefont {Chaves}},
  \bibinfo {author} {\bibfnamefont {F.~M.}\ \bibnamefont {Peeters}}, \ and\
  \bibinfo {author} {\bibfnamefont {G.~A.}\ \bibnamefont {Farias}},\ }\bibfield
   {title} {\enquote {\bibinfo {title} {Erratum: Simplified model for the
  energy levels of quantum rings in single layer and bilayer graphene [phys.
  rev. b 81, 045431 (2010)]},}\ }\href {\doibase 10.1103/PhysRevB.82.119906}
  {\bibfield  {journal} {\bibinfo  {journal} {Phys. Rev. B}\ }\textbf {\bibinfo
  {volume} {82}},\ \bibinfo {pages} {119906} (\bibinfo {year}
  {2010}{\natexlab{b}})}\BibitemShut {NoStop}%
\bibitem [{\citenamefont {Martin}\ \emph {et~al.}(2008)\citenamefont {Martin},
  \citenamefont {Blanter},\ and\ \citenamefont {Morpurgo}}]{Mar08}%
  \BibitemOpen
  \bibfield  {author} {\bibinfo {author} {\bibfnamefont {Ivar}\ \bibnamefont
  {Martin}}, \bibinfo {author} {\bibfnamefont {Ya.~M.}\ \bibnamefont
  {Blanter}}, \ and\ \bibinfo {author} {\bibfnamefont {A.~F.}\ \bibnamefont
  {Morpurgo}},\ }\bibfield  {title} {\enquote {\bibinfo {title} {Topological
  confinement in bilayer graphene},}\ }\href {\doibase
  10.1103/PhysRevLett.100.036804} {\bibfield  {journal} {\bibinfo  {journal}
  {Phys. Rev. Lett.}\ }\textbf {\bibinfo {volume} {100}},\ \bibinfo {pages}
  {036804} (\bibinfo {year} {2008})}\BibitemShut {NoStop}%
\bibitem [{\citenamefont {Zarenia}\ \emph {et~al.}(2011)\citenamefont
  {Zarenia}, \citenamefont {Pereira}, \citenamefont {Farias},\ and\
  \citenamefont {Peeters}}]{Zarenia11}%
  \BibitemOpen
  \bibfield  {author} {\bibinfo {author} {\bibfnamefont {M.}~\bibnamefont
  {Zarenia}}, \bibinfo {author} {\bibfnamefont {J.~M.}\ \bibnamefont
  {Pereira}}, \bibinfo {author} {\bibfnamefont {G.~A.}\ \bibnamefont {Farias}},
  \ and\ \bibinfo {author} {\bibfnamefont {F.~M.}\ \bibnamefont {Peeters}},\
  }\bibfield  {title} {\enquote {\bibinfo {title} {Chiral states in bilayer
  graphene: Magnetic field dependence and gap opening},}\ }\href {\doibase
  10.1103/PhysRevB.84.125451} {\bibfield  {journal} {\bibinfo  {journal} {Phys.
  Rev. B}\ }\textbf {\bibinfo {volume} {84}},\ \bibinfo {pages} {125451}
  (\bibinfo {year} {2011})}\BibitemShut {NoStop}%
\bibitem [{\citenamefont {Benchtaber}\ \emph
  {et~al.}(2021{\natexlab{a}})\citenamefont {Benchtaber}, \citenamefont
  {S\'anchez},\ and\ \citenamefont {Serra}}]{Ben21}%
  \BibitemOpen
  \bibfield  {author} {\bibinfo {author} {\bibfnamefont {Nassima}\ \bibnamefont
  {Benchtaber}}, \bibinfo {author} {\bibfnamefont {David}\ \bibnamefont
  {S\'anchez}}, \ and\ \bibinfo {author} {\bibfnamefont {Lloren{\c{c}}}\
  \bibnamefont {Serra}},\ }\bibfield  {title} {\enquote {\bibinfo {title}
  {Scattering of topological kink-antikink states in bilayer graphene
  structures},}\ }\href {\doibase 10.1103/PhysRevB.104.155303} {\bibfield
  {journal} {\bibinfo  {journal} {Phys. Rev. B}\ }\textbf {\bibinfo {volume}
  {104}},\ \bibinfo {pages} {155303} (\bibinfo {year}
  {2021}{\natexlab{a}})}\BibitemShut {NoStop}%
\bibitem [{\citenamefont {Xavier}\ \emph {et~al.}(2010)\citenamefont {Xavier},
  \citenamefont {Pereira}, \citenamefont {Chaves}, \citenamefont {Farias},\
  and\ \citenamefont {Peeters}}]{xavier10}%
  \BibitemOpen
  \bibfield  {author} {\bibinfo {author} {\bibfnamefont {L.~J.~P.}\
  \bibnamefont {Xavier}}, \bibinfo {author} {\bibfnamefont {J.~M.}\
  \bibnamefont {Pereira}}, \bibinfo {author} {\bibfnamefont {Andrey}\
  \bibnamefont {Chaves}}, \bibinfo {author} {\bibfnamefont {G.~A.}\
  \bibnamefont {Farias}}, \ and\ \bibinfo {author} {\bibfnamefont {F.~M.}\
  \bibnamefont {Peeters}},\ }\bibfield  {title} {\enquote {\bibinfo {title}
  {Topological confinement in graphene bilayer quantum rings},}\ }\href
  {\doibase 10.1063/1.3431618} {\bibfield  {journal} {\bibinfo  {journal}
  {Applied Physics Letters}\ }\textbf {\bibinfo {volume} {96}},\ \bibinfo
  {pages} {212108} (\bibinfo {year} {2010})}\BibitemShut {NoStop}%
\bibitem [{\citenamefont {Benchtaber}\ \emph
  {et~al.}(2021{\natexlab{b}})\citenamefont {Benchtaber}, \citenamefont
  {S{\'{a}}nchez},\ and\ \citenamefont {Serra}}]{Ben21b}%
  \BibitemOpen
  \bibfield  {author} {\bibinfo {author} {\bibfnamefont {Nassima}\ \bibnamefont
  {Benchtaber}}, \bibinfo {author} {\bibfnamefont {David}\ \bibnamefont
  {S{\'{a}}nchez}}, \ and\ \bibinfo {author} {\bibfnamefont {Lloren{\c{c}}}\
  \bibnamefont {Serra}},\ }\bibfield  {title} {\enquote {\bibinfo {title}
  {Geometry effects in topologically confined bilayer graphene loops},}\ }\href
  {\doibase 10.1088/1367-2630/ac434d} {\bibfield  {journal} {\bibinfo
  {journal} {New Journal of Physics}\ }\textbf {\bibinfo {volume} {24}},\
  \bibinfo {pages} {013001} (\bibinfo {year} {2021}{\natexlab{b}})}\BibitemShut
  {NoStop}%
\bibitem [{\citenamefont {Rozhkov}\ \emph {et~al.}(2016)\citenamefont
  {Rozhkov}, \citenamefont {Sboychakov}, \citenamefont {Rakhmanov},\ and\
  \citenamefont {Nori}}]{rozhkov16}%
  \BibitemOpen
  \bibfield  {author} {\bibinfo {author} {\bibfnamefont {A.V.}\ \bibnamefont
  {Rozhkov}}, \bibinfo {author} {\bibfnamefont {A.O.}\ \bibnamefont
  {Sboychakov}}, \bibinfo {author} {\bibfnamefont {A.L.}\ \bibnamefont
  {Rakhmanov}}, \ and\ \bibinfo {author} {\bibfnamefont {Franco}\ \bibnamefont
  {Nori}},\ }\bibfield  {title} {\enquote {\bibinfo {title} {Electronic
  properties of graphene-based bilayer systems},}\ }\href {\doibase
  https://doi.org/10.1016/j.physrep.2016.07.003} {\bibfield  {journal}
  {\bibinfo  {journal} {Physics Reports}\ }\textbf {\bibinfo {volume} {648}},\
  \bibinfo {pages} {1--104} (\bibinfo {year} {2016})},\ \bibinfo {note}
  {electronic properties of graphene-based bilayer systems}\BibitemShut
  {NoStop}%
\bibitem [{\citenamefont {Li}\ \emph {et~al.}(2016)\citenamefont {Li},
  \citenamefont {Wang}, \citenamefont {McFaul}, \citenamefont {Zern},
  \citenamefont {Ren}, \citenamefont {Watanabe}, \citenamefont {Taniguchi},
  \citenamefont {Qiao},\ and\ \citenamefont {Zhu}}]{Li16}%
  \BibitemOpen
  \bibfield  {author} {\bibinfo {author} {\bibfnamefont {Jing}\ \bibnamefont
  {Li}}, \bibinfo {author} {\bibfnamefont {Ke}~\bibnamefont {Wang}}, \bibinfo
  {author} {\bibfnamefont {Kenton~J.}\ \bibnamefont {McFaul}}, \bibinfo
  {author} {\bibfnamefont {Zachary}\ \bibnamefont {Zern}}, \bibinfo {author}
  {\bibfnamefont {Yafei}\ \bibnamefont {Ren}}, \bibinfo {author} {\bibfnamefont
  {Kenji}\ \bibnamefont {Watanabe}}, \bibinfo {author} {\bibfnamefont
  {Takashi}\ \bibnamefont {Taniguchi}}, \bibinfo {author} {\bibfnamefont
  {Zhenhua}\ \bibnamefont {Qiao}}, \ and\ \bibinfo {author} {\bibfnamefont
  {Jun}\ \bibnamefont {Zhu}},\ }\bibfield  {title} {\enquote {\bibinfo {title}
  {Gate-controlled topological conducting channels in bilayer graphene},}\
  }\href {\doibase 10.1038/nnano.2016.158} {\bibfield  {journal} {\bibinfo
  {journal} {Nature Nanotechnology}\ }\textbf {\bibinfo {volume} {11}},\
  \bibinfo {pages} {1060--1065} (\bibinfo {year} {2016})}\BibitemShut {NoStop}%
\end{thebibliography}%

\end{document}